\documentclass[aps,prl,twocolumn,superscriptaddress,preprintnumbers]{revtex4-1}%
\usepackage{epsf,epsfig}
\usepackage[utf8]{inputenc}
\usepackage{amssymb,amsmath,amsfonts}
\usepackage{color}
\usepackage{slashed}
\usepackage{tensor} 
\usepackage{braket}
\usepackage{array}
\usepackage{float}
\usepackage[colorlinks=true]{hyperref}  
\hypersetup{
    bookmarks=true,         
    unicode=false,          
    pdftoolbar=true,        
    pdfmenubar=true,        
    pdffitwindow=false,     
    pdfstartview={FitH},    
    colorlinks=true,       
    linkcolor=magenta, 
    citecolor=blue,        
    filecolor=magenta,      
    urlcolor=cyan           
} 

\newcommand{\bea}{\begin{eqnarray}}

\newcommand{\eea}{\end{eqnarray}}
\usepackage{lipsum} 
\newcommand{\ba}{\begin{eqnarray}}
\newcommand{\ea}{\end{eqnarray}}

\newcommand{\beq}{\begin{equation}}
\newcommand{\eeq}{\end{equation}}
\newcommand{\beqa}{\begin{eqnarray}}
\newcommand{\eeqa}{\end{eqnarray}}
\newcommand{\beqar}{\begin{eqnarray*}}
\newcommand{\eeqar}{\end{eqnarray*}}

\usepackage{color}

\newcommand{\dvtag}{\\ &}
\newcommand{\dvvtag}{\right.\\ &\left.}

\newcommand{\req}[1]{(\ref{#1})} 

\begin{document}

\title{Quasinormal modes of rotating black holes in higher-derivative gravity}

\author{Pablo A. Cano}
\email{pabloantonio.cano@kuleuven.be}
\affiliation{Institute for Theoretical Physics, KU Leuven. Celestijnenlaan 200D, B-3001 Leuven, Belgium}

\author{Kwinten Fransen}
\email{kfransen@ucsb.edu}
\affiliation{Department of Physics, University of California, Santa Barbara, CA93106, USA}

\author{Thomas Hertog}
\email{thomas.hertog@kuleuven.be}
\affiliation{Institute for Theoretical Physics, KU Leuven. Celestijnenlaan 200D, B-3001 Leuven, Belgium}

\author{Simon Maenaut}
\email{simon.maenaut@kuleuven.be}
\affiliation{Institute for Theoretical Physics, KU Leuven. Celestijnenlaan 200D, B-3001 Leuven, Belgium}

\date{\today}

\begin{abstract}
We compute the spectrum of linearized gravitational excitations of black holes with substantial angular momentum in the presence of higher-derivative corrections to general relativity. We do so perturbatively to leading order in the higher-derivative couplings and up to order fourteen in the black hole angular momentum. This allows us to accurately predict quasinormal mode frequencies of black holes with spins up to about $70\%$ of the extremal value.  
For some higher-derivative corrections, we find that sizable rotation enhances the frequency shifts by almost an order of magnitude relative to the static case.\\{ }\\{ }\\
\end{abstract} 

\maketitle

\textbf{Introduction:} 
Gravitational wave (GW) observations of binary black hole (BH) mergers \cite{LIGOScientific:2016aoc,LIGOScientific:2018mvr,LIGOScientific:2020ibl,LIGOScientific:2021usb,LIGOScientific:2021djp} probe highly relativistic aspects of the gravitational two-body problem and of the nature of BHs \cite{Yunes:2016jcc,LIGOScientific:2016lio,LIGOScientific:2018dkp,Berti:2018cxi,Berti:2018vdi,Barack:2018yly,LIGOScientific:2019fpa,LIGOScientific:2020tif,LIGOScientific:2021sio}. To go further and turn this novel observational window into a powerful tool to test general relativity (GR) requires a better understanding of the physically plausible corrections to GR and of their imprint on the patterns of GWs generated in BH collisions \cite{Sotiriou:2007zu, Yunes:2009ke,Yunes:2016jcc,Cardoso:2019rvt}. \\
\indent
Due to the substantial complexity, it is open for debate how to best approach this problem \cite{Giddings:2016tla,Hertog:2017vod,Witek:2018dmd,Okounkova:2019dfo,Bonilla:2022dyt,Maggio:2022hre}. However, a robust physical expectation is that a characteristic spectrum of complex frequencies determine the exponentially damped sinusoidal quasinormal modes (QNM) that govern the relaxation to the coalescence endstate \cite{Kokkotas:1999bd,Berti:2009kk}. These constitute the basic natural observables of the final ``ringdown'' stage of binary BH mergers.  \\
\indent
The spectrum of quasinormal modes frequencies is both experimentally accessible \cite{LIGOScientific:2018mvr,LIGOScientific:2020iuh,LIGOScientific:2020ibl,LIGOScientific:2021usb, Capano:2020dix, Capano:2021etf, Capano:2022zqm,Ma:2023cwe,Ma:2023vvr} and well-studied within GR \cite{regge1957stability, vishveshwara1970stability, zerilli1970effective, press1971long, moncrief1974gravitational,Newman:1961qr, price1972nonspherical,bardeen1973radiation,teukolsky1972rotating, teukolsky1973, teukolsky1974perturbations,Frolov:2017kze,Cook:2014cta,Stein:2019mop,Aminov:2020yma,Tanay:2022era,Gregori:2022xks,Fransen:2023eqj} and beyond \cite{Cardoso:2009pk,Molina:2010fb,Blazquez-Salcedo:2016enn,Blazquez-Salcedo:2017txk,Tattersall:2018nve,Konoplya:2020bxa,Moura:2021eln,Moura:2021nuh,deRham:2020ejn,Cardoso:2018ptl,McManus:2019ulj,Pani:2012bp,Pani:2013pma,Pierini:2021jxd,Wagle:2021tam,Srivastava:2021imr,Cano:2021myl,Pierini:2022eim}. Yet, as a result of significant technical difficulties, the beyond-GR results have so far been restricted to static or slowly rotating BHs whereas most merger remnants have substantial  angular momentum \cite{LIGOScientific:2018mvr,LIGOScientific:2020ibl,LIGOScientific:2021usb,LIGOScientific:2021djp,LIGOScientific:2021psn}.\\
\indent
In this Letter, we report on the very first explicit computations of QNM frequencies for rotating BHs in a general class of beyond-GR theories, for BH spins of prime astrophysical interest. Our results are the culmination of a large body of work \cite{Cano:2019ore,Cano:2020cao,Cano:2021myl,Li:2022pcy,Hussain:2022ins,Cano:2023tmv}, but specifically deliver on the promise of the method established and validated at low spin in \cite{Cano:2023tmv}. We refer to this work for the details of the computational method we employ. Here we concentrate on the derivation of the values of the QNM frequency shifts for BHs with relatively large spin. First, however, we briefly review the effective field theory (EFT) extension of GR that sets our theoretical framework. Throughout, we work in geometric units $G=c=1$.\\

\textbf{Effective field theory of gravity:}
The most general EFT extension of GR to eight derivatives is given by the action \footnote{Observe that this action does not contain quadratic terms, since these do not affect Ricci flat solutions and hence are irrelevant for EFT.}
\begin{equation}\label{eq:EFT}
\begin{aligned}
S=\frac{1}{16\pi}\int& d^4x\sqrt{|g|}\bigg[R+\ell^4\left(\lambda_{\rm ev}\mathcal{R}^3+\lambda_{\rm odd}\tilde{\mathcal{R}}^3\right)\\
+&\ell^6\left(\epsilon_{1}\mathcal{C}^2+\epsilon_{2}\tilde{\mathcal{C}}^2+\epsilon_{3}\mathcal{C}\tilde{\mathcal{C}}\right)+\mathcal{O}(\ell^8) \bigg]\, ,
\end{aligned}
\end{equation}
with higher-derivative curvature scalars
\begin{align*}
\mathcal{R}^3&=\tensor{R}{_{\mu\nu }^{\rho\sigma}}\tensor{R}{_{\rho\sigma }^{\delta\gamma }}\tensor{R}{_{\delta\gamma }^{\mu\nu }}\, ,& \tilde{\mathcal{R}}^3&=\tensor{R}{_{\mu\nu }^{\rho\sigma}}\tensor{R}{_{\rho\sigma }^{\delta\gamma }} \tensor{\tilde R}{_{\delta\gamma }^{\mu\nu }}\, ,\\
\mathcal{C}&=R_{\mu\nu\rho\sigma} R^{\mu\nu\rho\sigma}\, ,&
\tilde{\mathcal{C}}&=R_{\mu\nu\rho\sigma} \tilde{R}^{\mu\nu\rho\sigma}\, , 
\end{align*}
where $\tilde{R}_{\mu\nu\rho\sigma}=\frac{1}{2}\epsilon_{\mu\nu\alpha\beta}\tensor{R}{^{\alpha\beta}_{\rho\sigma}}$. 
The length scale $\ell$ is related to the cutoff of the EFT, $\ell\sim 1/\Lambda_{\rm cutoff}$,  and the coefficients $\lambda_{\rm ev, odd}$, $\epsilon_{1,2,3}$ are dimensionless. Using the scale set by a BH of mass $M$, we also introduce the dimensionless couplings 
\begin{equation}
\alpha_{\rm ev}=\frac{\ell^4\lambda_{\rm ev}}{M^4}\, ,\quad \alpha_{\rm odd}=\frac{\ell^4\lambda_{\rm odd}}{M^4}\, ,\quad \alpha_{i}=\frac{\ell^6\epsilon_{i}}{M^6}\, .
\end{equation}
These couplings characterize the size of the relative corrections to GR, which thus become larger for smaller BHs. 

Throughout our analysis we assume $|\alpha_{\rm q} |\ll 1$, where ${\rm q} \in \left\lbrace {\rm ev, \rm odd, 1, 2, 3}\right\rbrace$, so that we are well within the EFT regime and can work perturbatively in these couplings. The couplings can be further constrained by considerations about symmetries or causality \cite{Gruzinov:2006ie,Endlich:2017tqa,Chen:2021bvg,deRham:2021bll}. However, here we do not impose those or other constraints but rather present a general analysis. 
Finally, we note that the action \eqref{eq:EFT} only contains terms that are unambiguous under field redefinitions. Since QNM frequencies are invariant under field redefinitions, this theory captures the most general corrections to the QNM frequencies of BHs in vacuum GR up to eight derivatives.\\

\textbf{Rotating black holes:}  The rotating BH solutions of \eqref{eq:EFT} can be described by the ansatz \cite{Cano:2019ore}
\begin{align}\label{eq:ansatz}
    ds^2 =& -\left(1-\frac{2M r}{\Sigma}-H_1\right)dt^2\\\notag
    &
    -(1+ H_2)\frac{4a M r \sin^2\theta}{\Sigma}dtd\phi
    \\\notag
    &+\left(1+H_3\right)\Sigma\left(\frac{dr^2}{\Delta}+d\theta^2\right)\\\notag
    &+(1+H_4)\left(r^2 + a^2+\frac{2a^2Mr\sin^2\theta}{\Sigma}\right)\sin^2\theta d\phi^2\, , 
\end{align}
where we have four functions $H_j(r,\theta)$ deforming the Kerr metric in Boyer-Lindquist coordinates, and where $ \Sigma=r^2+a^2\cos^2\theta$, $\Delta=r^2-2Mr+a^2$, and $M$ and $a$ represent the BH mass and specific angular momentum, respectively.  We only consider first-order corrections in the $\alpha_{\rm q}$ couplings, and hence we have $H_j=\sum_{\rm q} \alpha_{\rm q}H_{j,\rm q}+\mathcal{O}(\alpha_{\rm q}^2)$. The functions $H_{j,\rm q}$ satisfy a system of partial linear differential equations with no known closed-form analytic solutions. Nonetheless it is possible to obtain an analytic solution expressed as a power series in the dimensionless spin $\chi=a/M$ \cite{Cano:2019ore}.  
This expansion takes the form 
\begin{equation}
\label{eq:H-exp}
    H_{j,\rm q} = \sum_{n=0}^{\infty}\chi^n  \sum_{p=0}^{n} \sum_{k=0}^{k_\text{max}(n)}H_{j, \rm q}^{(n,p,k)}\left(\frac{M}{r}\right)^k\cos^p\theta \, ,
\end{equation}
where each term is a finite polynomial in $\cos(\theta)$ and $1/r$ with coefficients $H_{j, \rm q}^{(n,p,k)}$ that we determine analytically. In order to compute BH observables with a given accuracy it suffices to include only a finite number of terms in the above expansions \cite{Cano:2023qqm}. Below we use expansions up to order  $\mathcal{O}(\chi^{14})$, which we show to be sufficient to quantify the corrections to the QNMs for BHs of spins $\chi\sim 0.7$ to within a few percent. 

\textbf{Corrected radial Teukolsky equations:}
Gravitational perturbations of rotating BHs \req{eq:ansatz} in the theory \req{eq:EFT} can be characterized in terms of the perturbed Weyl scalars $\delta\Psi_{0}$, $\delta\Psi_{4}$, $\delta\Psi_{0}^{*}$ and $\delta\Psi_{4}^{*}$ \footnote{The conjugate variables $\delta\Psi_{0,4}^{*}$ are considered independent variables, since we naturally work with a complex metric perturbation. Thus, the symbol $*$ denotes Newman-Penrose (NP) conjugation, obtained upon the exchange $m_{\mu}\leftrightarrow \bar{m}_{\mu}$ in the NP frame. For a complex metric perturbation, this is no longer equivalent to complex conjugation.}, which satisfy modified Teukolsky equations \cite{Li:2022pcy,Hussain:2022ins,Cano:2023tmv}. 
As shown in \cite{Cano:2023tmv},  one can reduce these general Teukolsky equations  to a system of four radial equations for four master variables $R^{lm}_{s}(r)$, $R^{*lm}_{s}(r)$ by decomposing the Weyl scalars into spin-weighted spheroidal harmonics. Here, $(l,m)$ are the angular mode harmonic numbers and $s=\pm 2$ is the spin weight of each variable.  At first order in the higher-derivative corrections, these variables satisfy decoupled modified radial Teukolsky equations, which read
\begin{equation}\label{eq:correctedradial}
\Delta^{-s+1}\frac{d}{dr}\left[\Delta^{s+1}\frac{dR_{s}}{dr}\right]+\left(V_s+ \delta V_s\right) R_{s}=0\, ,
\end{equation}   where we have suppressed the $lm$ labels for clarity.
In addition we have two equations for the conjugate variables $R_{s}^{*}$ that take the same form, except with a different $\delta V_s$ (but the same $V_s$).   

In \eqref{eq:correctedradial}, $V_s$ is the usual Teukolsky potential --- that we review in the Supplemental Material ---  and the effect of the higher-derivative terms is encoded in the correction to the potential, $\delta V_{s}$, which in general involves a linear combination of all the corrections, $\delta V_{s}=\sum_{\rm q}\alpha_{\rm q}\delta V_{s,{\rm q}}+\mathcal{O}(\alpha_{\rm q}^2)$. Interestingly, we find that for the theories under consideration it can always be expressed in the simple form
\begin{equation}\label{eq:deltaVs}
\delta V_{s}=\frac{A_{-2}}{r^2}+\sum_{n=0}^{4}A_{n}r^n\, .
\end{equation}
Thus, the entire correction to the Teukolsky equation is determined by the six coefficients $A_{n}$. These coefficients depend on the quantum numbers $s$, $l$ and $m$, and they are functions of $M$,  $a$, the frequency $\omega$ and two additional sets of parameters $q_{\pm 2}$ and $C_{\pm 2}$.
The former, $q_{\pm 2}$, are related to the polarization of the perturbation and hence they are physical. On the other hand, $C_{\pm 2}$ are Starobinsky-Teukolsky (ST) constants that relate the Hertz potentials with spins $s=\pm 2$ in the metric perturbation. They roughly represent different ways in which one can reconstruct the metric perturbation from the Teukolsky variables \cite{Pound:2021qin,Dolan:2021ijg}. 
Therefore, they arise from a redundancy in our description of the perturbations and hence physical quantities like QNM frequencies should be independent of them. We exploit this to test our methods and to estimate the accuracy of our results. 

Our approach allows us to obtain the coefficients $A_n$ analytically as a power series in the BH's dimensionless spin,
\begin{equation}
A_n=\sum_{k=0}^{\infty} A_{n,(k)}\chi^k\, .
\end{equation}
For the present work, we obtain the expansion of $\delta V_s$ to order $\mathcal{O}(\chi^{14})$ for the $(l,m)=(2,2)$ modes and to order $\mathcal{O}(\chi^{12})$ for the $(3,3)$ modes of all theories in \req{eq:EFT}. 
Additional details on the form of the $A_n$ coefficients are provided in the Supplemental Material. 

\textbf{Quasinormal modes:}
Quasinormal modes are the solutions of the equations \req{eq:correctedradial} with outgoing boundary conditions at infinity and at the horizon. Thus they represent the relaxation of the BH after a local perturbation. 
This provides an eigenvalue problem for the frequency $\omega$, whose solutions are the QNM frequencies. An additional feature in the case of higher-derivative gravity theories is that the Teukolsky equations depend on the polarization of the perturbation, and this must be determined alongside with the frequency. Therefore, one must find the values of $\omega$ and $q_{\pm 2}$ by solving the different Teukolsky equations simultaneously. Concretely, given some values of the polarization parameters $q_{\pm 2}$, we first solve the four different radial equations independently. We therefore obtain four (possible) values for the QNM frequencies. We denote these respectively by $\omega_{s}(q_{\pm 2})$ and $\omega_{s}^{*}(q_{\pm 2})$. 
Next we demand that
\begin{equation}\label{eq:equalomegas}
\omega_{+2}(q_{\pm 2})=\omega_{+2}^{*}(q_{\pm 2})=\omega_{-2}(q_{\pm 2})=\omega_{-2}^{*}(q_{\pm 2})\, .
\end{equation}
This provides a system of equations for the polarization parameters $q_{\pm 2}$.

We now consider parity-preserving and parity-breaking theories as two separate cases, and then move on to general theories containing both types of corrections. 

In parity-preserving theories --- like the extensions of GR with the terms $\mathcal{R}^3$, $\mathcal{C}^2$ and $\tilde{\mathcal{C}}^2$ in \req{eq:EFT} ---  the modes of odd and even parity decouple.  This allows us to fix the polarization parameters to $q_{+2}=q_{-2}=\pm 1$, where the ``$+$'' sign corresponds to polar perturbations and the ``$-$'' sign to axial ones. With each of these choices, the equations for the conjugate variables $R_{s}^{*}$ become identical to those of $R_{s}$ (\textit{i.e.}, $\delta V_{s}^{*}=\delta V_{s}$). The QNM frequencies $\omega$ are then obtained by solving any of the radial equations, either $s=2$ or $s=-2$. Crucially, both equations give the same result and the latter is independent of the ST constants $C_{\pm 2}$ \cite{Cano:2023tmv}.  Since our approach computes QNM frequencies that are linearly corrected with respect to the Kerr values, we write
\begin{equation}\label{eq:deltaomegaeven}
\omega^{\pm}_{\rm parity\, \, preserving}=\omega_{\rm Kerr}+\delta\omega^{\pm} \, ,
\end{equation}  
where the label $\pm$ refers to the polarization. Generically we observe that $\delta\omega^{+}\neq \delta\omega^{-}$, implying that the isospectrality of axial and polar modes is broken. 

When the theory contains higher-derivative terms that break parity --- in our case the terms $\tilde{\mathcal{R}}^3$ and $\mathcal{C}\tilde{\mathcal{C}}$ in \req{eq:EFT} ---  modes of odd and even parity intermingle. As a consequence, the parameters $q_{\pm 2}$, which determine the relative weight of each mode, must be determined by actually solving \req{eq:equalomegas}. 
We find that these equations indeed possess two solutions for the parameters $q_{\pm 2}$, giving rise to two QNMs of different polarization. More details on the structure of these equations are in the Supplemental Material. The associated frequencies take the form 
\begin{equation}\label{eq:deltaomegaodd}
\omega^{\pm}_{\rm parity\,\, breaking}=\omega_{\rm Kerr}\pm \delta\omega_{\rm break}\, ,
\end{equation}  
\textit{i.e.} the shift in one of the frequencies is the opposite of the other. This behavior is characteristic of parity-violating corrections \cite{Cano:2021myl}.
We also remark that, even though we still use the labels $\pm$, these no longer refer to the parity of the mode, as neither of these modes has a definite parity. 

A general higher-derivative theory may contain both parity-preserving and parity-violating higher-derivative terms. In that case, the total correction to the QNM frequencies is not simply given by a linear superposition of the corrections. Instead, one must again solve the equations \req{eq:equalomegas} and include all terms in these. Doing so yields the following ``combination rule'',
\begin{equation}\label{eq:combinationrule}
\delta\omega_{\rm total}^{\pm}=\frac{\delta\omega^{+}+\delta\omega^{-}}{2}\pm  \sqrt{\frac{\left(\delta\omega^{+}-\delta\omega^{-}\right)^2}{4}+\delta\omega_{\rm break}^2}\, ,
\end{equation}
where $\delta\omega^{\pm}$ are the shifts in the frequency if only parity-preserving corrections were present --- as in \req{eq:deltaomegaeven} --- while $\delta\omega_{\rm break}$ is the shift coming from the parity-breaking corrections if only those were present --- as in \req{eq:deltaomegaodd}. 

\textbf{Computation of quasinormal mode frequencies:}
Following the previous discussion, QNM frequencies are labeled by the numbers $(l,m,n)$ --- $n$ being the overtone number --- and by the polarization type ``$\pm$''.  In order to find these frequencies, we can either solve the $s=+2$ or $s=-2$ equations, since they give the same results --- this was already checked in \cite{Cano:2023tmv}. 
However,  they do not perform equally well under the spin expansion on which our method relies. Specifically, the $s=-2$ equations show greater stability and faster convergence in the slow-rotation expansion than their $s=+2$ counterparts. Therefore, 
we consider the $s=-2$ equations in what follows. 

\begin{figure}[t!]
	\centering
	\includegraphics[width=0.48\textwidth]{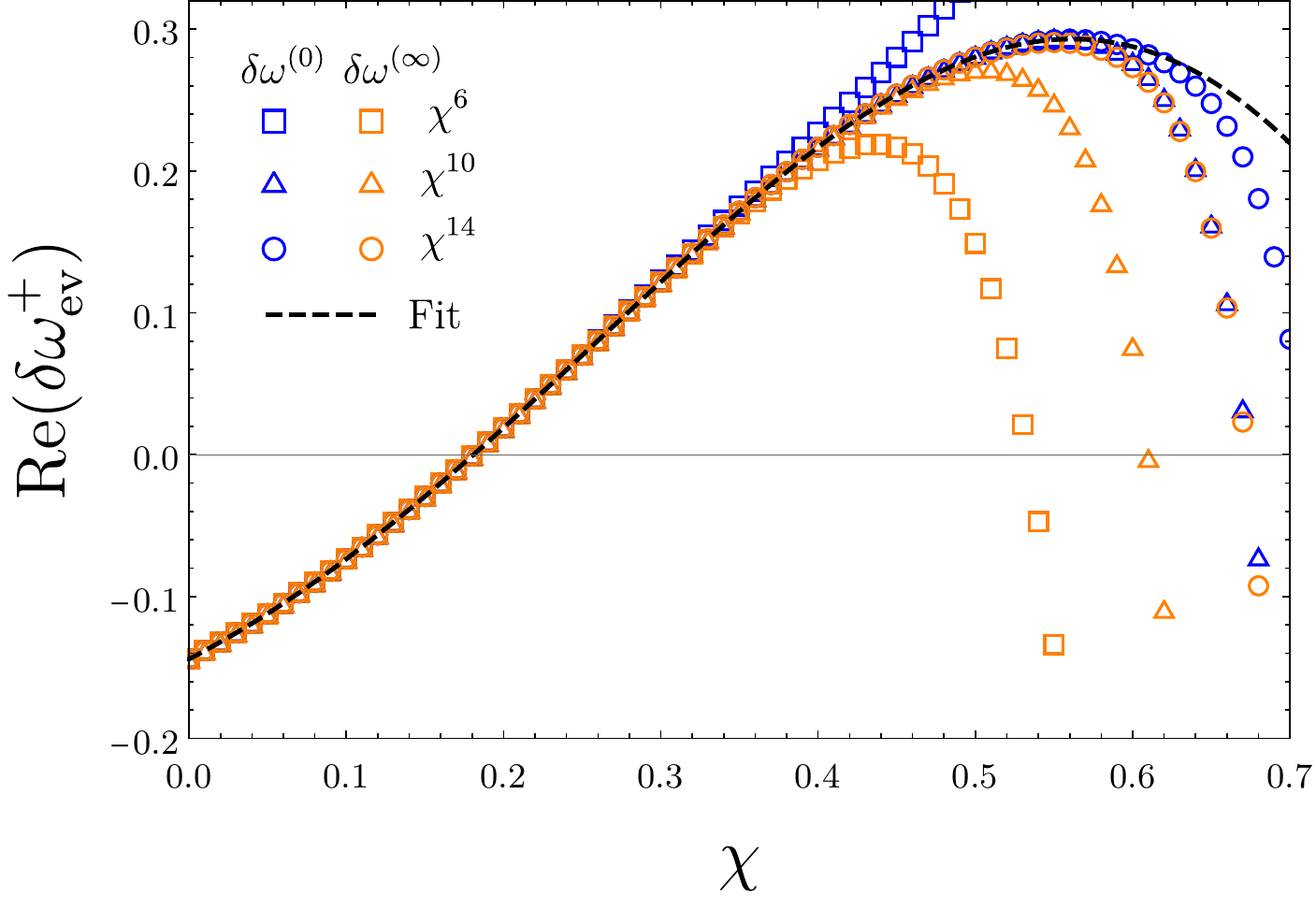}
	\caption{Shift in the $(2,2,0)$ polar QNM frequency due to the cubic term $\mathcal{R}^3$. We show the real part of $\delta\omega_{\rm ev}^{+}$ as a function of the angular momentum computed at several orders in the spin expansion. Blue and orange points correspond to the two different estimates, $\delta\omega^{(0)}$ and $\delta\omega^{(\infty)}$, calculated with $C_{-2}=0$ and $C_{-2}\rightarrow\infty$, respectively. The dashed black line represents a weighted polynomial fit of order 12.}
	\label{fig:convergencel2}
\end{figure}

\begin{figure*}[t!]
			\centering
			\includegraphics[width=0.49\textwidth]{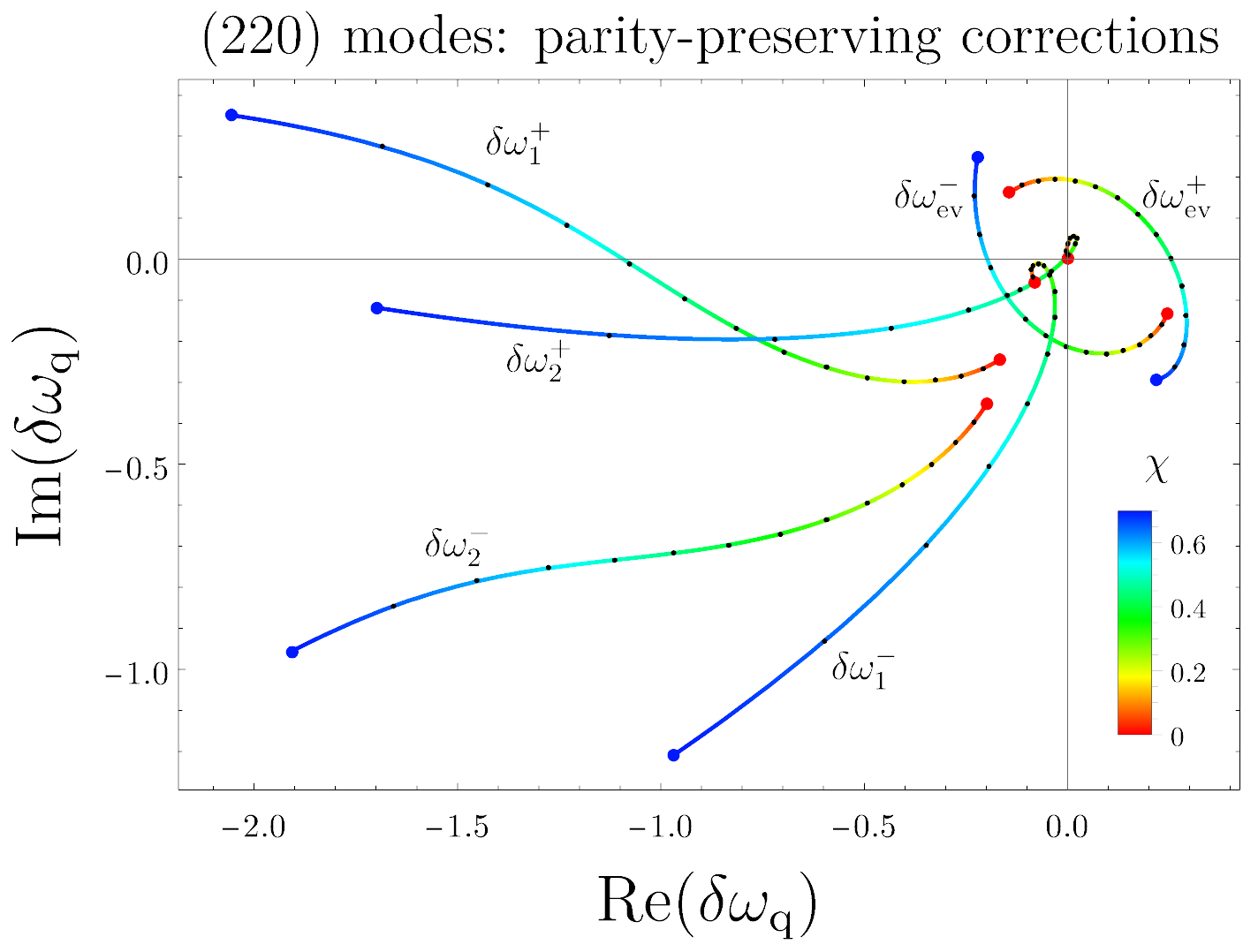}
			\includegraphics[width=0.49\textwidth]{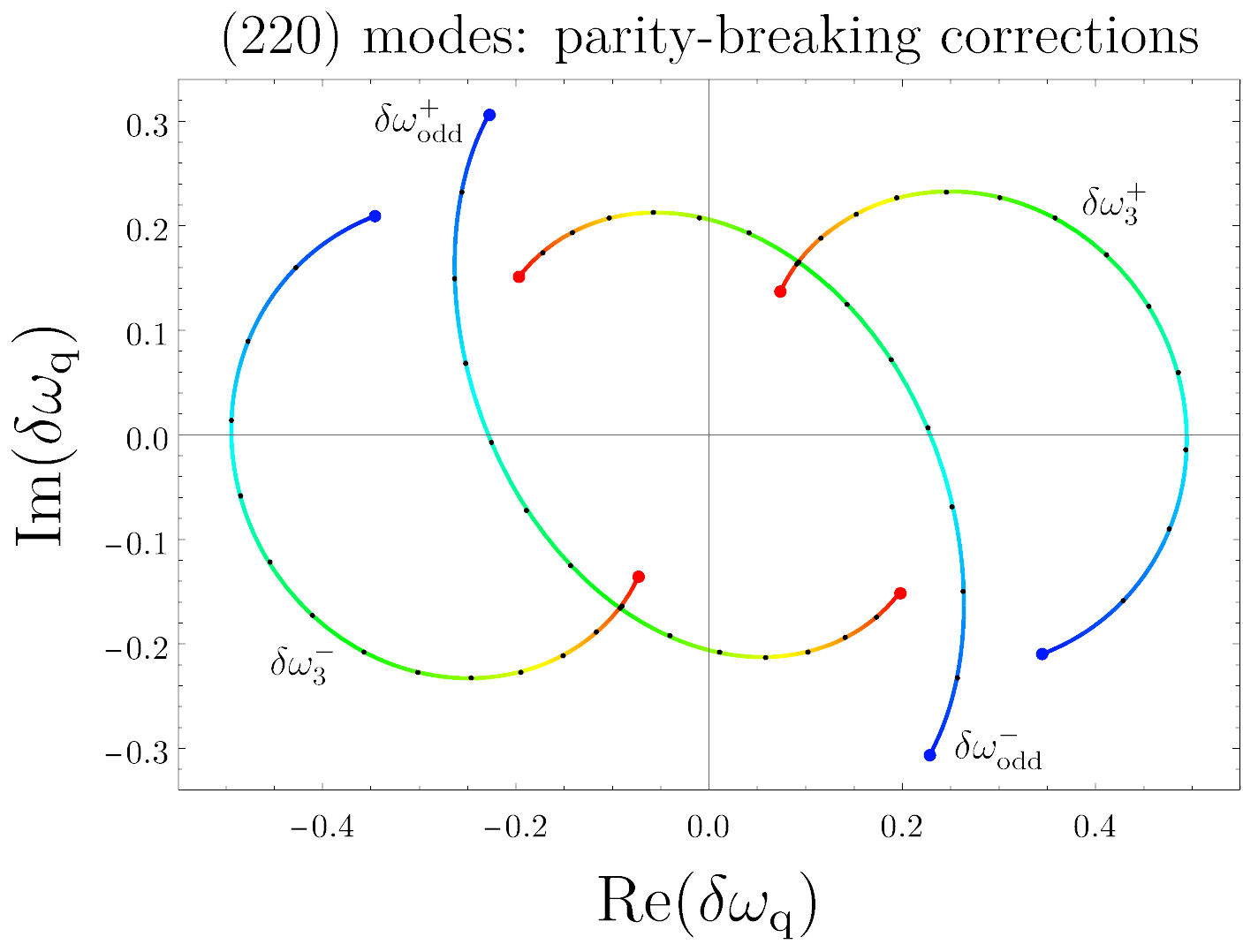}
			\includegraphics[width=0.49\textwidth]{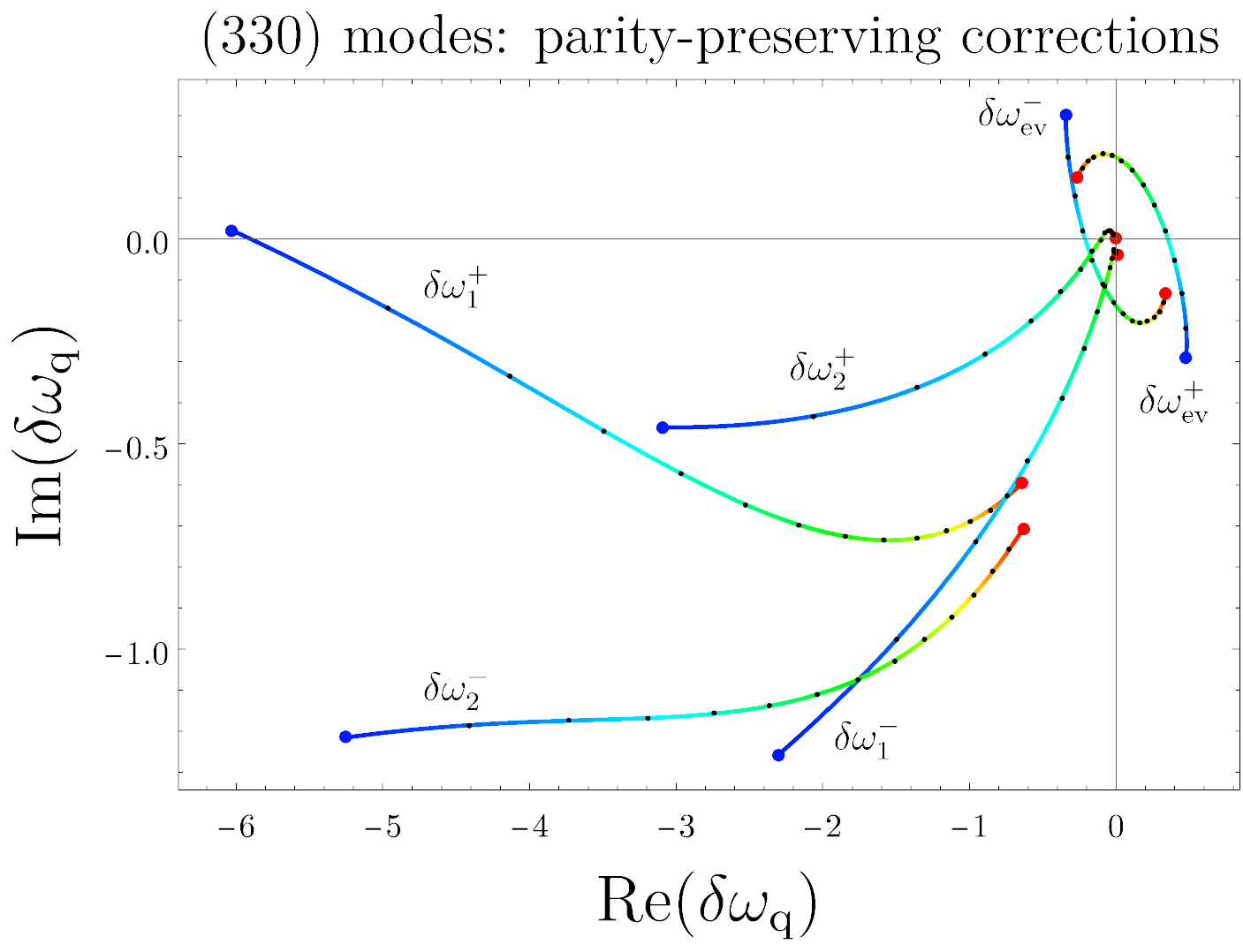}
			\includegraphics[width=0.49\textwidth]{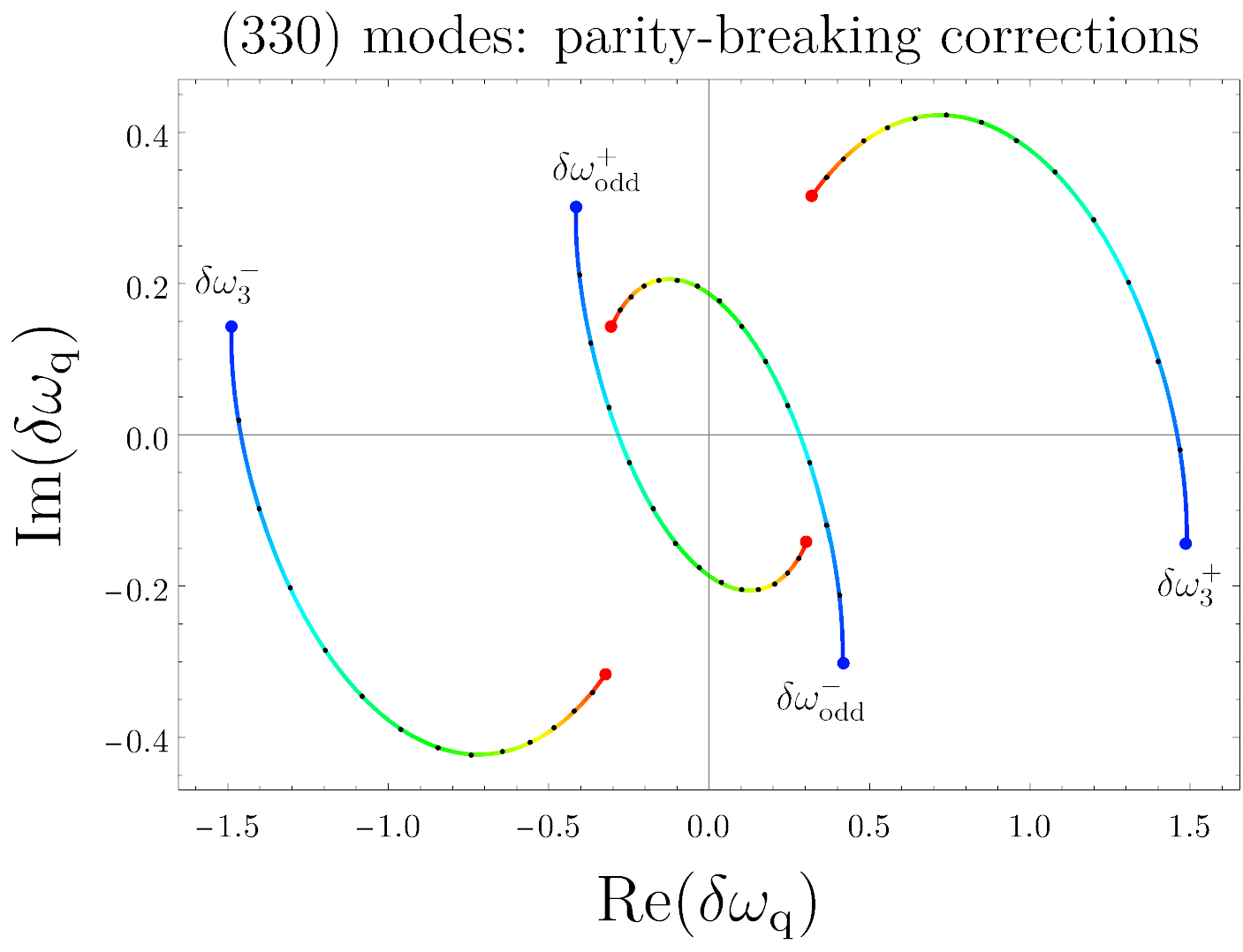}
			\caption{Corrections to the $(2,2,0)$  and $(3,3,0)$ QNM frequencies. We show the running of the coefficients $\delta\omega_{\rm q}$ (defined in \req{eq:deltaomegaqdef}) in the complex plane as a function of the angular momentum $\chi$. The red dots correspond to static BHs and the blue dots to rotating BHs of spin $\chi=0.7$. The intermediate black dots represent an increase in spin of $0.05$. Left: frequency shifts for parity-preserving theories. We observe that the shifts of different polarization $\delta\omega^{\pm}$ are independent. Right: parity-violating theories. In this case, the shift for one polarization is the opposite of the other, $\delta\omega^{+}=-\delta\omega^{-}$.}
			\label{fig:complexplane22}
     \end{figure*} 
     
Now, these equations depend on the undetermined ST constant $C_{-2}$, which is useful to estimate the accuracy of our results. The QNM frequencies should be independent of this constant, but this is only exactly true when we include the full series in the spin. Since we are truncating the series expansion at some finite order $n$, the result is independent of $C_{-2}$ only up to terms of order $\mathcal{O}(\chi^{n+1})$. 
Thus, in order to estimate the error in our results on account of the truncation of the spin expansion, we compute the QNM frequencies for $C_{-2}=0$ and $C_{-2}\rightarrow\infty$. The difference between the two estimates, denoted by $\delta\omega^{(0)}$ and $\delta\omega^{(\infty)}$, respectively, informs us about the accuracy we are achieving. 

We illustrate this in Fig.~\ref{fig:convergencel2}, where we show, for the two different choices of $C_{-2}$, the shift in the $(l,m,n)=(2,2,0)$ polar mode due to the cubic correction $\mathcal{R}^3$ computed at different orders in the spin expansion. 
In this plot, and in all of the subsequent results, we report the value of the correction $\delta\omega_{\rm q}$ to the frequency defined by
\begin{equation}\label{eq:deltaomegaqdef}
\omega=\omega_{\rm Kerr}+\frac{\alpha_{\rm q}}{M}\delta\omega_{\rm q}\, ,
\end{equation}
for each of the corrections. 
Fig.~\ref{fig:convergencel2} clearly shows that, as we increase the order of the expansion, $\delta\omega^{(0)}$ and $\delta\omega^{(\infty)}$ remain close for larger values of the angular momentum, hence indicating that the spin expansion is convergent. We observe a similar behavior in all the other QNM frequencies we have computed. 
We also see that, in the domain where our results are accurate, $\delta\omega_{\rm q}$ is a very smooth function of $\chi$.  This allows us to extrapolate the results to slightly larger spins by fitting the numerical results to a polynomial (black dashed line in Fig.~\ref{fig:convergencel2}), 
\begin{equation}
\delta\omega_{\rm fit}(\chi)=\sum_{n=0}^{N}c_n \chi^n\, ,
\end{equation} 
where we weight each data point by the inverse of the variance between the two estimates, $w=(\delta\omega^{(0)}-\delta\omega^{(\infty)})^{-2}$. 

We have performed this analysis and computed these polynomial fits for the $(2,2,0)$ and $(3,3,0)$ modes for all the theories in \req{eq:EFT}.  The coefficients for these fits are provided in the Supplemental Material, and they represent the key output of our work. 

These polynomials provide an approximation that is practically indistinguishable from the numerical results in the region where these are accurate, which, depending on the case, goes up to $\chi\sim 0.5-0.6$. However, the fits allow us to extrapolate the results to somewhat larger spins, up to the desired value of $\chi=0.7$ and even beyond. One can check the accuracy of this extrapolation by comparing the fits obtained from a lower-order spin expansion. As we show in the Supplemental Material, this leads to the conclusion that our results for $\delta\omega_{\rm q} $ have an accuracy of at least $5\%$ for $\chi=0.7$, and they should also provide a reasonable estimation even for $\chi\sim 0.8$.

We offer a visualization of our results in Fig.~\ref{fig:complexplane22}, where we show the trajectories in the complex plane of the correction coefficients $\delta\omega_{\rm q}$ for the $(2,2,0)$ and $(3,3,0)$ modes for the two different polarizations --- which are no longer isospectral. This figure demonstrates that the behavior of these coefficients is highly non-linear in $\chi$ and that the shifts in the QNM frequencies can completely change, in both magnitude and sign, for BHs of different spins.

\textbf{Discussion:}

We have obtained the QNM frequency shifts of rotating BHs in a general EFT extension of vacuum GR.  Specifically, we have presented explicit results for the $(2,2,0)$ and $(3,3,0)$ modes up to spins $\chi\sim 0.7$. In this regime of angular momentum --- in the relevant range for post-coalescence BH ringdowns ---  our results for the frequency shifts have an accuracy of around $5\%$, which is more than enough for observational purposes. For instance, if the frequency shift is a 2\% relative to the Kerr value (consistent with current data, but potentially observable with future GW detectors), then the error in the total frequency is merely a $0.1\%$. This is smaller than the expected experimental uncertainties for future experiments. 
Therefore, our results allow for precision tests of higher-derivative corrections in ringdown signals and should be valuable input for the phenomenological spectroscopy of BHs with gravitational waves in the coming years and decades.

Looking ahead, it will be important to compute the spectral shifts for other harmonics as well as for overtones, and to extend our results to even higher BH spins. The latter could be achieved by increasing the order of the spin expansion. Although entirely algorithmic, this poses a significant computational challenge that likely requires further optimization to resolve. Also a complementary, complete numerical approach could prove useful to study the regime of even higher spins, if only as an independent cross-check of the results we have presented. 

From a phenomenological viewpoint it would be of great interest to identify configurations in which, perhaps in certain models, the corrections to the QNM spectrum become exceptionally large. In this respect we note that the near-extremal limit $\chi\rightarrow 1$ may amplify the effects of higher-derivative corrections \cite{Horowitz:2023xyl}. Although our current results are only valid far from extremality, they do hint indeed at a significant growth of the shifts with the spin. For instance, in the quartic theories the shifts rapidly increase along the negative real axis for increasing angular momentum, changing by an order of magnitude from static BHs to $\chi\sim 0.7$ (see Fig.~\ref{fig:complexplane22}). Moreover, causality constraints impose $\epsilon_{1,2}\ge 0$ \cite{Gruzinov:2006ie,Endlich:2017tqa} and therefore preclude a cancellation between terms. This suggests that in due course, detailed GW observations of BH mergers involving large spins may well provide an arena for some of the strongest tests of beyond-GR physics in BH spectroscopy.

\vskip1mm
\noindent
\textbf{Acknowledgements}
\vskip1mm

\noindent
S.M. would like to thank Marina David for help with Mathematica. The work of P.A.C. is supported by a postdoctoral fellowship from the Research Foundation - Flanders (FWO grant 12ZH121N). K.F. is supported by the Heising-Simons Foundation grant \#2021-2819. T.H. acknowledges support from the PRODEX grant LISA - BEL (PEA 4000131558), the FWO Research Project G0H9318N and the inter-universitary project iBOF/21/084. This work made use of the Mathematica BH perturbation toolkit \cite{BHPToolkit}, and took inspiration from the open source Python QNM package by Leo C. Stein \cite{Stein:2019mop} and the Julia code written by Asad Hussain \cite{Hussain:2022ins}.

\onecolumngrid  \vspace{1cm} 
\begin{center}  
{\Large\bf Appendix} 
\end{center} 
\appendix

 \section{Quasinormal mode frequencies: polynomial fits}
Here we provide our results for the correction to the QNM frequencies, $\delta\omega_{\rm q}(\chi)$, defined by (12) in the main text.  We present $\delta\omega_{\rm q}(\chi)$ in the form of polynomial fits, that we obtain as follows.
First, given a higher-derivative theory, we numerically compute the shifts in the QNM frequencies $\delta\omega^{(0)}$ and $\delta\omega^{(\infty)}$ (obtained by setting $C_{-2}=0$ or $C_{-2}\rightarrow\infty$) for  $-0.7<\chi<0.7$ with steps of $0.01$.  Then, for each $\chi$ we take the average value of both estimations, $\delta\omega_{\rm av.}=(\delta\omega^{(0)}+\delta\omega^{(\infty)})/2$ and we perform a weighted polynomial fit,
\begin{equation}\label{eq:polynomialfit}
\delta\omega_{\rm fit}(\chi)=\sum_{n=0}^{N}c_n \chi^n\, ,
\end{equation}  
where as weight for each data point we use the inverse of the variance between the two estimates, $w=(\delta\omega^{(0)}-\delta\omega^{(\infty)})^{-2}$. This creates a bias towards the data points of smaller $\chi$, since they have much smaller errors. 
In order to determine the optimum order of the polynomial, $N$, we perform these fits with increasing values of $N$ and look at the sum of weighted square residuals. This decreases exponentially with $N$ until a certain $N_{\rm optimum}$, when the error saturates and no longer decreases. By choosing $N=N_{\rm optimum}$ we make sure that we are not overfitting the data and that the result is meaningful. We find that, with the numeric results obtained at order $\mathcal{O}(\chi^{14})$ for the $(2,2,0)$ modes, we typically have $N_{\rm optimum}=12$. For the $(3,3,0)$ modes, computed from a $\mathcal{O}(\chi^{12})$ expansion of Teukolsky equation, we also find $N_{\rm optimum}=12$, because the results for higher $l$ seem to be more precise.
 
The polynomial coefficients are shown in tables \ref{table:qnm-shift-cubic-fit-220}, \ref{table:qnm-shift-quartic-fit-220},  \ref{table:qnm-shift-cubic-fit-330},  and \ref{table:qnm-shift-quartic-fit-330}. Importantly, we observe that the results up to linear order in the spin approximately match with those of \cite{Cano:2021myl}. We remark however that the agreement does not need to be exact since, strictly speaking, our polynomial fits are not the same as a Taylor expansion. 

One may also try to perform a Pad\'e resummation instead of polynomial fits \cite{Pierini:2022eim}. However, we found that the Pad\'e approximants may give rise to singularities due to the higher-order expansion, and we generally observed that they do not provide a better convergence than our polynomials. 

Finally, in tables \ref{table3} and \ref{table4} we provide an estimate of the accuracy of these polynomials when extrapolated at $\chi = 0.7$. We do this by comparing the estimate from these polynomials with the one obtained from a lower-order spin expansion (of order $12$ and $10$ for the $(2,2,0)$ and $(3,3,0)$ modes, respectively), fitted with a $10\textsuperscript{th}$ order polynomial.

\section{Numerical method}

The numerical method used to compute the shifts in the QNM frequencies is largely based on the approach of \cite{Hussain:2022ins}, in~the spirit of finding the spectral shifts of a perturbed Hermitian operator. We implement a Leaver solver with the Cook-Zalutskiy spectral approach to the angular sector \cite{Cook:2014cta} to find the Kerr QNM frequencies and radial functions. As we are interested in the first-order effect of the corrections we can treat each term of $\delta V_s$ (from (5) in the main text) in (\ref{eq:deltaVCq}) separately. We can thus compute each term of (\ref{eq:deltaomegaCq}) (see below) individually. In the analysis we only use the the $s = -2$ equation and for parity preserving corrections we set $q_s = \pm 1$, thereby only requiring $\delta V_s$, while for the parity breaking corrections we, a priori, have to solve for $q_s$ requiring both $\delta V_s$ and $\delta V_s^*$ for solutions in the limits of $C_{-2}$ going to zero and to infinity.
We also cross-checked the results by employing a direct numerical integration of Teukolsky equations with a shooting method to find the QNM frequencies, finding good agreement.

\bgroup
\def\arraystretch{1.24}
\setlength{\tabcolsep}{4pt}
\begin{table}[H]
	\centering
	\begin{tabular}{|c||c|c|c|}
	\hline
 \text{} & $ \delta\omega^+_{\text{ev}} $ & $ \delta\omega^-_{\text{ev}} $ & $ \delta\omega^+_{\text{odd}} $ \\
 \hline \hline
 $ c_0 $  & $ -0.144+0.162 i $ & $ +0.246-0.132 i $ & $ +0.197-0.151 i $ \\ \hline
 $ c_1 $  & $ +0.572+0.411 i $ & $ -0.256-0.590 i $ & $ -0.422-0.512 i $ \\ \hline
 $ c_2 $  & $ +1.389-0.858 i $ & $ -1.407+0.253 i $ & $ -1.431+0.561 i $ \\ \hline
 $ c_3 $  & $ -0.361-2.254 i $ & $ -0.422+2.007 i $ & $ -0.059+2.183 i $ \\ \hline
 $ c_4 $  & $ -2.378-0.321 i $ & $ +1.718+1.052 i $ & $ +2.065+0.759 i $ \\ \hline
 $ c_5 $  & $ -0.931+1.869 i $ & $ +1.065-0.976 i $ & $ +1.045-1.376 i $ \\ \hline
 $ c_6 $  & $ +0.640+1.103 i $ & $ -0.286-0.621 i $ & $ -0.420-0.843 i $ \\ \hline
 $ c_7 $  & $ +0.305+0.419 i $ & $ -0.040-0.075 i $ & $ -0.138-0.243 i $ \\ \hline
 $ c_8 $  & $ +0.279+0.521 i $ & $ -0.022-0.276 i $ & $ -0.128-0.407 i $ \\ \hline
 $ c_9 $  & $ +0.301+0.249 i $ & $ -0.133-0.108 i $ & $ -0.207-0.196 i $ \\ \hline
 $ c_{10} $  & $ +0.112+0.212 i $ & $ -0.007-0.105 i $ & $ -0.058-0.182 i $ \\ \hline
 $ c_{11} $  & $ +0.118+0.164 i $ & $ -0.063-0.119 i $ & $ -0.099-0.174 i $ \\ \hline
 $ c_{12} $  & $ +0.003+0.021 i $ & $ -0.006-0.008 i $ & $ -0.018-0.040 i $ \\ \hline
	\end{tabular}
	\caption{Best-fit coefficients for the polynomial fits \req{eq:polynomialfit} of the shifts in the $(2,2,0)$ QNM frequencies for cubic higher derivative corrections.}
	\label{table:qnm-shift-cubic-fit-220}
\end{table}

\begin{table}[H]
	\centering
	\begin{tabular}{|c||c|c|c|c|c|}
	\hline
 \text{} & $ \delta\omega^+_1 $ & $ \delta\omega^-_1 $ & $ \delta\omega^+_2 $ & $ \delta\omega^-_2 $ & $ \delta\omega^+_3 $ \\
 \hline \hline
 $ c_0 $  & $ -0.166-0.244 i $ & $ -0.079-0.058 i $ & $ +0.000+0.000 i $ & $ -0.199-0.352 i $ & $ +0.074+0.136 i $ \\ \hline
 $ c_1 $  & $ -0.715-0.451 i $ & $ -0.209+0.321 i $ & $ -0.100+0.415 i $ & $ -0.556-0.885 i $ & $ +0.252+0.527 i $ \\ \hline
 $ c_2 $  & $ -2.145+0.176 i $ & $ +1.082+0.303 i $ & $ +1.104+0.019 i $ & $ -1.982-0.955 i $ & $ +1.625+0.280 i $ \\ \hline
 $ c_3 $  & $ -1.751+3.219 i $ & $ +1.434-2.793 i $ & $ +0.165-3.041 i $ & $ -2.587+1.879 i $ & $ +1.552-2.790 i $ \\ \hline
 $ c_4 $  & $ +3.221+2.870 i $ & $ -3.187-4.046 i $ & $ -5.216-2.409 i $ & $ +1.814+3.167 i $ & $ -3.366-3.239 i $ \\ \hline
 $ c_5 $  & $ +3.064-3.563 i $ & $ -5.012+1.128 i $ & $ -5.350+3.968 i $ & $ +3.585-2.121 i $ & $ -4.353+2.633 i $ \\ \hline
 $ c_6 $  & $ -3.236-3.820 i $ & $ -0.339+2.750 i $ & $ +0.515+4.329 i $ & $ -1.323-3.754 i $ & $ +1.089+3.691 i $ \\ \hline
 $ c_7 $  & $ -3.529+0.407 i $ & $ +0.422-0.587 i $ & $ +0.645+0.091 i $ & $ -2.120-0.378 i $ & $ +1.623-0.076 i $ \\ \hline
 $ c_8 $  & $ -1.918+0.081 i $ & $ -1.261-0.346 i $ & $ -1.293+0.370 i $ & $ -0.607-0.666 i $ & $ -0.042+0.219 i $ \\ \hline
 $ c_9 $  & $ -2.732+0.117 i $ & $ -0.463-0.053 i $ & $ -0.602+0.338 i $ & $ -1.492-0.863 i $ & $ +0.780+0.328 i $ \\ \hline
 $ c_{10} $  & $ -2.197+0.707 i $ & $ -0.940-0.749 i $ & $ -1.414-0.366 i $ & $ -1.162-0.224 i $ & $ +0.253-0.341 i $ \\ \hline
 $ c_{11} $  & $ -2.284+0.321 i $ & $ -1.175-0.320 i $ & $ -1.669+0.067 i $ & $ -1.228-0.610 i $ & $ +0.174+0.075 i $ \\ \hline
 $ c_{12} $  & $ -1.918+0.754 i $ & $ -0.639-0.645 i $ & $ -1.081-0.443 i $ & $ -1.248-0.035 i $ & $ +0.383-0.375 i $ \\ \hline
	\end{tabular}
	\caption{Best-fit coefficients for the polynomial fits \req{eq:polynomialfit} of the shifts in the $(2,2,0)$ QNM frequencies for quartic higher derivative corrections.}
	\label{table:qnm-shift-quartic-fit-220}
\end{table}

\begin{table}[H]
	\centering
	\begin{tabular}{|c||c|c|c|}
	\hline
 \text{} & $ \delta\omega^+_{\text{ev}} $ & $ \delta\omega^-_{\text{ev}} $ & $ \delta\omega^+_{\text{odd}} $ \\
 \hline \hline
 $ c_0 $  & $ -0.265+0.150 i $ & $ +0.343-0.134 i $ & $ +0.304-0.142 i $ \\ \hline
 $ c_1 $  & $ +0.554+0.438 i $ & $ -0.347-0.463 i $ & $ -0.451-0.452 i $ \\ \hline
 $ c_2 $  & $ +1.440-0.425 i $ & $ -1.316+0.274 i $ & $ -1.381+0.349 i $ \\ \hline
 $ c_3 $  & $ +0.602-1.635 i $ & $ -0.718+1.490 i $ & $ -0.666+1.566 i $ \\ \hline
 $ c_4 $  & $ -1.037-1.127 i $ & $ +0.785+1.186 i $ & $ +0.906+1.165 i $ \\ \hline
 $ c_5 $  & $ -1.253+0.348 i $ & $ +1.056-0.111 i $ & $ +1.153-0.219 i $ \\ \hline
 $ c_6 $  & $ -0.513+0.845 i $ & $ +0.415-0.570 i $ & $ +0.465-0.698 i $ \\ \hline
 $ c_7 $  & $ -0.222+0.662 i $ & $ +0.172-0.407 i $ & $ +0.200-0.527 i $ \\ \hline
 $ c_8 $  & $ -0.183+0.608 i $ & $ +0.162-0.385 i $ & $ +0.176-0.491 i $ \\ \hline
 $ c_9 $  & $ -0.088+0.529 i $ & $ +0.082-0.350 i $ & $ +0.089-0.437 i $ \\ \hline
 $ c_{10} $  & $ -0.077+0.415 i $ & $ +0.076-0.276 i $ & $ +0.081-0.348 i $ \\ \hline
 $ c_{11} $  & $ -0.097+0.388 i $ & $ +0.104-0.264 i $ & $ +0.104-0.325 i $ \\ \hline
 $ c_{12} $  & $ -0.065+0.281 i $ & $ +0.081-0.191 i $ & $ +0.071-0.228 i $ \\ \hline
	\end{tabular}
	\caption{Best-fit coefficients for the polynomial fits \req{eq:polynomialfit} of the shifts  in the $(3,3,0)$ QNM frequencies for cubic higher derivative corrections.}
	\label{table:qnm-shift-cubic-fit-330}
\end{table}

\begin{table}[H]
	\centering
	\begin{tabular}{|c||c|c|c|c|c|}
	\hline
 \text{} & $ \delta\omega^+_1 $ & $ \delta\omega^-_1 $ & $ \delta\omega^+_2 $ & $ \delta\omega^-_2 $ & $ \delta\omega^+_3 $ \\
 \hline \hline
 $ c_0 $  & $ -0.641-0.595 i $ & $ +0.006-0.040 i $ & $ +0.000-0.000 i $ & $ -0.634-0.706 i $ & $ +0.321+0.316 i $ \\ \hline
 $ c_1 $  & $ -1.832-0.663 i $ & $ -0.126+0.135 i $ & $ -0.246+0.173 i $ & $ -1.768-1.000 i $ & $ +0.810+0.494 i $ \\ \hline
 $ c_2 $  & $ -3.079-0.075 i $ & $ +0.196-0.024 i $ & $ +0.122+0.046 i $ & $ -2.747-0.677 i $ & $ +1.544+0.195 i $ \\ \hline
 $ c_3 $  & $ -3.680+1.525 i $ & $ +0.110-1.620 i $ & $ -0.136-1.482 i $ & $ -3.265+0.766 i $ & $ +1.743-1.351 i $ \\ \hline
 $ c_4 $  & $ -1.869+2.890 i $ & $ -2.141-2.950 i $ & $ -2.871-2.494 i $ & $ -1.652+2.225 i $ & $ -0.357-2.650 i $ \\ \hline
 $ c_5 $  & $ +0.157+1.042 i $ & $ -4.510-1.371 i $ & $ -5.388-0.429 i $ & $ +0.454+0.716 i $ & $ -2.617-0.905 i $ \\ \hline
 $ c_6 $  & $ -1.166-1.324 i $ & $ -3.917+0.866 i $ & $ -4.678+2.040 i $ & $ -0.513-1.520 i $ & $ -1.725+1.422 i $ \\ \hline
 $ c_7 $  & $ -3.014-0.988 i $ & $ -2.505+0.750 i $ & $ -3.186+1.883 i $ & $ -2.114-1.287 i $ & $ -0.141+1.216 i $ \\ \hline
 $ c_8 $  & $ -3.215-0.335 i $ & $ -2.585+0.176 i $ & $ -3.306+1.239 i $ & $ -2.199-0.714 i $ & $ -0.121+0.609 i $ \\ \hline
 $ c_9 $  & $ -3.425-0.351 i $ & $ -2.603+0.325 i $ & $ -3.350+1.316 i $ & $ -2.316-0.793 i $ & $ -0.054+0.698 i $ \\ \hline
 $ c_{10} $  & $ -3.822-0.088 i $ & $ -2.495+0.143 i $ & $ -3.287+1.043 i $ & $ -2.628-0.610 i $ & $ +0.158+0.484 i $ \\ \hline
 $ c_{11} $  & $ -3.628+0.034 i $ & $ -2.878+0.057 i $ & $ -3.520+0.861 i $ & $ -2.463-0.577 i $ & $ -0.137+0.278 i $ \\ \hline
 $ c_{12} $  & $ -2.033-0.169 i $ & $ -2.236+0.235 i $ & $ -2.478+0.724 i $ & $ -1.354-0.596 i $ & $ -0.432+0.182 i $ \\ \hline
	\end{tabular}
	\caption{Best-fit coefficients for the polynomial fits \req{eq:polynomialfit} of the shifts in the $(3,3,0)$ QNM frequencies for quartic higher derivative corrections.}
	\label{table:qnm-shift-quartic-fit-330}
\end{table}
\egroup

\bgroup
\def\arraystretch{1.24}
\setlength{\tabcolsep}{4pt}
\begin{table}[H]
	\centering
	\begin{tabular}{|c|c|c|c|}
	\hline
$\delta\omega$ & $\mathcal{O}(\chi^{12})$&  $\mathcal{O}(\chi^{14})$ & Diff. $\%$ \\
\hline \hline
$ \delta\omega^+_{\text{ev}} $ & $ +0.217-0.297 i $ & $ +0.220-0.293 i $ & $ 1.23 $ \\ \hline
$ \delta\omega^-_{\text{ev}} $ & $ -0.220+0.253 i $ & $ -0.221+0.251 i $ & $ 0.88 $ \\ \hline
$ \delta\omega^+_{\text{odd}} $ & $ -0.226+0.310 i $ & $ -0.228+0.306 i $ & $ 1.25 $ \\ \hline
$ \delta\omega^+_1 $ & $ -1.979+0.331 i $ & $ -2.056+0.351 i $ & $ 3.82 $ \\ \hline
$ \delta\omega^-_1 $ & $ -0.932-1.191 i $ & $ -0.967-1.208 i $ & $ 2.49 $ \\ \hline
$ \delta\omega^+_2 $ & $ -1.645-0.113 i $ & $ -1.697-0.118 i $ & $ 3.09 $ \\ \hline
$ \delta\omega^-_2 $ & $ -1.866-0.944 i $ & $ -1.908-0.957 i $ & $ 2.08 $ \\ \hline
$ \delta\omega^+_3 $ & $ +0.337-0.204 i $ & $ +0.345-0.209 i $ & $ 2.34 $ \\ \hline
	\end{tabular}
	\caption{Shift in the $(2,2,0)$ QNM frequencies for $\chi=0.7$. We show the prediction based on the fitting polynomials obtained from the numerical results at order $\chi^{12}$ and order $\chi^{14}$ as well as the relative difference between the two estimations. 
	} \label{table3}
\end{table}

\begin{table}[H]
	\centering
	\begin{tabular}{|c|c|c|c|}
	\hline
$\delta\omega$ & $\mathcal{O}(\chi^{10})$&  $\mathcal{O}(\chi^{12})$ & Diff. $\%$ \\
\hline \hline
$ \delta\omega^+_{\text{ev}} $ & $ +0.481-0.308 i $ & $ +0.478-0.291 i $ & $ 3.11 $ \\ \hline
$ \delta\omega^-_{\text{ev}} $ & $ -0.350+0.315 i $ & $ -0.345+0.301 i $ & $ 3.20 $ \\ \hline
$ \delta\omega^+_{\text{odd}} $ & $ -0.419+0.317 i $ & $ -0.415+0.301 i $ & $ 3.09 $ \\ \hline
$ \delta\omega^+_1 $ & $ -5.909+0.030 i $ & $ -6.033+0.021 i $ & $ 2.05 $ \\ \hline
$ \delta\omega^-_1 $ & $ -2.186-1.269 i $ & $ -2.300-1.257 i $ & $ 4.38 $ \\ \hline
$ \delta\omega^+_2 $ & $ -2.949-0.500 i $ & $ -3.089-0.460 i $ & $ 4.67 $ \\ \hline
$ \delta\omega^-_2 $ & $ -5.180-1.184 i $ & $ -5.254-1.216 i $ & $ 1.50 $ \\ \hline
$ \delta\omega^+_3 $ & $ +1.510-0.158 i $ & $ +1.489-0.143 i $ & $ 1.75 $ \\ \hline
	\end{tabular}
	\caption{Shift in the $(3,3,0)$ QNM frequencies for $\chi=0.7$. We show the prediction based on the fitting polynomials obtained from the numerical results at order $\chi^{10}$ and order $\chi^{12}$ as well as the relative difference between the two estimations. 
	} \label{table4}
\end{table}
\egroup

\section{The corrected radial Teukolsky equations}
Our starting point is the corrected radial Teukolsky equations as given by \cite{Cano:2023tmv}. These are a set of four equations, for $R_{\pm 2}$ and $R_{\pm }^{*}$, and read
\begin{equation}\label{radialeqs}
\begin{aligned}
P_{s}\mathfrak{D}_{s}^2R_{s}-\left[ f_{s} R_{s}+g_{s}\frac{dR_{s}}{dr}\right]=&0\, ,\\
P_{s}^{*}\mathfrak{D}_{s}^2R_{s}^{*}-\left[ f_{s}^{*}R_{s}^{*}+g_{s}^{*}\frac{dR_{s}^{*}}{dr}\right]=&0\, ,
\end{aligned}
\end{equation}
where 
\begin{equation}
\mathfrak{D}_{s}^2=\Delta^{-s+1}\frac{d}{dr}\left[\Delta^{s+1}\frac{d}{dr}\right]+V_{s} 
\end{equation}
is the Teukolsky operator, where the Teukolsky potential $V_s$ is given by 
\begin{equation}
\begin{aligned}
V_s&=(am)^2+\omega^2 \left(a^2+r^2\right)^2-4 a m M r \omega+i s \left(2 a m (r-M)-2 M \omega \left(r^2-a^2\right)\right)+\Delta\left(-a^2 \omega^2+s-B_{lm}+2 i r s \omega\right)\, ,
\end{aligned}
\end{equation}  
and $B_{lm}$ are the angular separation constants for the usual spin-weighted spheroidal harmonics, in the conventions of \cite{Cano:2023tmv}, such that $B_{lm}(a\omega=0)=l(l+1)-s^2$.  
In addition, $P_{s}$, $P_{s}^{*}$ are constants that we specify below and $f_{s}$ and $g_{s}$ are functions of $r$ proportional to the higher-derivative couplings which take the form
\begin{equation}
\begin{aligned}
f_{s}(r)&=M^2\sum_{n=0}^{N}\left(\frac{r}{M}\right)^{4-n}f_{s,n}\, ,\\
g_{s}(r)&=M(r-r_{+})\sum_{n=0}^{N}\left(\frac{r}{M}\right)^{3-n}g_{s,n}\, ,
\end{aligned}
\end{equation}
and analogously for the conjugates $f_{s}^{*}$,  $g_{s}^{*}$. Here $f_{s,n}$, $g_{s,n}$ are coefficients taking the form of a power series in the spin $\chi$ and $N$ increases with the order of this expansion, so that $N\rightarrow\infty$ for the full series but it takes a finite value when we truncate it at a finite order.

Now, by performing a change of variables of the form
\begin{equation}\label{eq:changeofvariables}
\begin{aligned}
R_{s}&\rightarrow R_{s}+\left(\alpha_{s}R_{s}+\beta_{s}\frac{dR_{s}}{dr}\right)\, ,\\
R_{s}^{*}&\rightarrow R_{s}^{*}+ \left(\alpha_{s}^{*}R_{s}^{*}+\beta_{s}^{*}\frac{dR_{s}^{*}}{dr}\right)\, ,
\end{aligned}
\end{equation}
where the functions $\alpha_{s}$ and $\beta_{s}$ are linear in the higher-derivative couplings, 
we can rewrite \req{radialeqs} as \eqref{eq:correctedradial} in the main text, 
\begin{equation}\label{eq:correctedradialapp} 
\begin{aligned}
\Delta^{-s+1}\frac{d}{dr}\left[\Delta^{s+1}\frac{dR_{s}}{dr}\right]+\left(V_s+ \delta V_s\right) R_{s}=&0\, ,\\
\Delta^{-s+1}\frac{d}{dr}\left[\Delta^{s+1}\frac{dR_{s}^{*}}{dr}\right]+\left(V_s+ \delta V_s^{*}\right) R_{s}^{*}=&0\, .
\end{aligned}
\end{equation}  
To this end, it suffices to consider the $\alpha_{s}$ and $\beta_{s}$ functions as polynomials in $1/r$ with appropriate coefficients.  This also allows us to put $\delta V_s$ in the form \eqref{eq:deltaVs} (in the main text), where, interestingly, the coefficient of $r^{-2}$ cannot be set to zero. It must also be noted that in the implementation of the change of variables \req{eq:changeofvariables} to derive \req{eq:correctedradialapp} we only keep terms up to first order in the higher-derivative couplings and we make use of the uncorrected Teukolsky equation for $R_{s}$ in the terms proportional to the corrections in order to reduce the number of derivatives.

\section{Structure of the potential and the QNM equations}

The potential $V_s$, and hence the coefficients $A_n$, have the following dependence on the $q_{\pm 2}$ and $C_{\pm 2}$ constants \cite{Cano:2023tmv}: 
  \begin{equation}\label{eq:deltaVCq}
 \begin{aligned}
 \delta V_{s}&=\frac{1}{P_{s}}\left[\delta V_{s}^{(1)}+q_{s}\delta V_{s}^{(2)}+C_{s}\delta V_{s}^{(3)}+C_{s}q_{-s}\delta V_{s}^{(4)}\right]\, ,\\
 \delta V_{s}^{*}&=\frac{1}{q_{s}P_{s}^{*}}\left[\delta V_{s}^{*(1)}+q_{s}\delta V_{s}^{*(2)}+C_s \delta V_{s}^{*(3)}+q_{-s}C_{s}\delta V_{s}^{*(4)}\right]\, ,
 \end{aligned}
 \end{equation}

where  $\delta V_{s}^{(j)}$,   $\delta V_{s}^{*(j)}$, are functions of $r$ that depend on $a$, $M$ and $\omega$.  We also have
\begin{equation}\label{eq:Pconstants}
\begin{aligned}
P_{+2}&=\frac{1}{2}+i K_{1} q_{+2}-i  K_{2} C_{+2} q_{-2} \, ,\\
P_{-2}&=\frac{1}{2}-i K_{1} q_{-2}+i  \frac{K_{2} C_{-2} q_{2}}{16} \, ,\\
P_{+2}^{*}&=\frac{1}{2}+\frac{i K_1}{q_{+2}}-\frac{i K_{2} C_{+2}}{q_{+2}}\, ,\\
P_{-2}^{*}&=\frac{1}{2}-\frac{i K_1}{q_{-2}}+\frac{i K_{2} C_{-2}}{16 q_{-2}}\, ,
\end{aligned}
\end{equation}
where 
\begin{equation}
K_{1}= \frac{D_{2}}{24 M \omega}\, ,\quad K_2= \frac{D_{2}^2+144 M^2 \omega^2}{6M\omega}\, ,
\end{equation}
and 
\begin{equation}\label{D2value}
\begin{aligned}
D_{2}=\Big[&\left(8+6 B_{lm}+B_{lm}^2\right)^2-8 \left(-8+B_{lm}^2 (4+B_{lm})\right) m \gamma +4 \left(8-2 B_{lm}-B_{lm}^2+B_{lm}^3\dvvtag
+2 (-2+B_{lm}) (4+3 B_{lm})
m^2\right) \gamma ^2-8 m \left(8-12 B_{lm}+3 B_{lm}^2+4 (-2+B_{lm}) m^2\right) \gamma ^3\dvtag
+2 \left(42-22 B_{lm}+3 B_{lm}^2+8 (-11+3
B_{lm}) m^2+8 m^4\right) \gamma ^4\dvtag-8 m \left(3 B_{lm}+4 \left(-4+m^2\right)\right) \gamma ^5
+4 \left(-7+B_{lm}+6
m^2\right) \gamma ^6-8 m \gamma ^7+\gamma ^8\Big]^{1/2}\, ,
\end{aligned}
\end{equation}
where $\gamma=a\omega$. The value of $\omega$ in these expressions can be set to the Kerr value, since we are only interested in the leading correction. 

Now, the corrections to the frequencies computed from each equation (denoted by $\delta \omega_{s}$ and $\delta \omega_{s}^{*}$) inherit the same dependence on the $q_{\pm 2}$ and $C_{\pm 2}$ constants as the correction to the potential, and hence in general we have
\begin{equation}\label{eq:deltaomegaCq}
\begin{aligned}
\delta \omega_{s}&=\frac{1}{P_{s}}\left[\delta\omega_{s}^{(1)}+q_{s}\delta\omega_{s}^{(2)}+C_{s}\delta\omega_{s}^{(3)}+C_{s}q_{-s}\delta\omega_{s}^{(4)}\right]\, ,\\
\delta \omega_{s}^{*}&=\frac{1}{q_{s}P_{s}^{*}}\left[\delta\omega_{s}^{*(1)}+q_{s}\delta\omega_{s}^{*(2)}+C_s \delta\omega_{s}^{*(3)}+q_{-s}C_{s}\delta\omega_{s}^{*(4)}\right]\, ,
\end{aligned}
\end{equation}
where $\delta\omega_{s}^{(j)}$, $\delta\omega_{s}^{*(j)}$ are coefficients that we find numerically by solving the QNM equations. Not all of these coefficients are independent, though, since these equations possess some structure.

In the case of parity-preserving corrections we observe $\delta\omega_{s}^{(j)}=\delta\omega_{s}^{*(j)}$ (because the conjugate equations are identical to the non-conjugate ones), while the coefficients $\delta\omega_{s}^{(j)}$ satisfy 

\begin{equation}\label{eq:relationsparityeven}\begin{aligned}
\left\{\delta\omega_{+2}^{(j)}\right\}_{j=1}^{4}&=\left\{\frac{16 K_1 \delta \omega ^{\text{(3)}}_{-2}-8 i \delta \omega ^{\text{(4)}}_{-2}}{K_2},\, \frac{16 K_1\delta \omega ^{\text{(4)}}_{-2}-8i \delta \omega ^{\text{(3)}}_{-2}}{K_2},\,-16 \delta \omega ^{\text{(3)}}_{-2},\,-16 \delta \omega ^{\text{(4)}}_{-2}\right\}\, ,\\
\left\{\delta\omega_{-2}^{(j)}\right\}_{j=1}^{4}&=\left\{-\frac{16 K_1 \delta \omega ^{\text{(3)}}_{-2}+8i \delta \omega ^{\text{(4)}}_{-2}}{K_2},\,-\frac{16 K_1 \delta \omega ^{\text{(4)}}_{-2}+8i \delta \omega ^{\text{(3)}}_{-2}}{K_2},\, \delta \omega ^{\text{(3)}}_{-2},\,\delta \omega ^{\text{(4)}}_{-2}\right\}\, .
\end{aligned}
\end{equation}

These relations imply that the total shift in the QNM frequencies is given by 
\begin{equation}\label{deltaomegaevenapp}
\delta\omega_{+2}\Big|_{q_{+2}=q_{-2}=\pm1}=\delta\omega_{-2}\Big|_{q_{+2}=q_{-2}=\pm1}=-\frac{16 i \left(\pm \delta \omega ^{\text{(3)}}{}_{-2}+\delta \omega ^{\text{(4)}}{}_{-2}\right)}{K_2}\equiv \delta\omega^{\pm}\, ,
\end{equation}
so that QNMs indeed correspond to modes of definite parity ($q_{+2}=q_{-2}=\pm 1$) and the result is independent of the ST constants $C_{\pm 2}$.  

The coefficients $\delta\omega_{s}^{(j)}$, $\delta\omega_{s}^{*(j)}$ for parity-violating theories have a more complicated structure, but we find that all of them can again be written in terms of only two parameters, for instance $-\delta \omega ^{\text{(4)}}_{-2}$ and $-\delta \omega ^{\text{(3)}}_{-2}$

\begin{equation}\label{eq:relationsparityodd}\begin{aligned}
\left\{\delta\omega_{+2}^{(j)}\right\}_{j=1}^{4}&=\left\{\frac{16 K_1 \delta \omega ^{\text{(3)}}_{-2}+8 i \delta \omega ^{\text{(4)}}_{-2}}{K_2},\,\frac{16 K_1 \delta \omega ^{\text{(4)}}_{-2}+8 i \delta \omega ^{\text{(3)}}_{-2}}{K_2},\,-16 \delta \omega ^{\text{(3)}}_{-2},\,16 \delta \omega ^{\text{(4)}}_{-2}\right\}\, ,\\
\left\{\delta\omega_{+2}^{*(j)}\right\}_{j=1}^{4}&=\left\{-\frac{16 K_1 \delta \omega ^{\text{(4)}}_{-2}+8i \delta \omega ^{\text{(3)}}_{-2}}{K_2},\,-\frac{16 K_1 \delta \omega ^{\text{(3)}}_{-2}+8i \delta \omega ^{\text{(4)}}_{-2}}{K_2},\,-16 \delta \omega ^{\text{(4)}}_{-2},\,16 \delta \omega ^{\text{(3)}}_{-2}\right\}\, ,\\
\left\{\delta\omega_{-2}^{(j)}\right\}_{j=1}^{4}&=\left\{-\frac{16 K_1 \delta \omega ^{\text{(3)}}_{-2}+8i \delta \omega ^{\text{(4)}}_{-2}}{K_2},\,\frac{16 K_1 \delta \omega ^{\text{(4)}}_{-2}+8 i \delta \omega ^{\text{(3)}}_{-2}}{K_2},\,\delta \omega ^{\text{(3)}}_{-2},\,\delta \omega ^{\text{(4)}}_{-2}\right\}\, ,\\
\left\{\delta\omega_{-2}^{*(j)}\right\}_{j=1}^{4}&=\left\{-\frac{16 K_1 \delta \omega ^{\text{(4)}}_{-2}+8i \delta \omega ^{\text{(3)}}_{-2})}{K_2},\,\frac{16 K_1 \delta \omega ^{\text{(3)}}_{-2}+8 i \delta \omega ^{\text{(4)}}_{-2}}{K_2},\,-\delta \omega ^{\text{(4)}}_{-2},\,-\delta \omega ^{\text{(3)}}_{-2}\right\}\, ,\\
\end{aligned}
\end{equation}
These relations imply that the system of equations

\begin{equation}\label{eq:equationsomegaapp}
\delta\omega_{+2}=\delta\omega_{+2}^{*}=\delta\omega_{-2}=\delta\omega_{-2}^{*}
\end{equation}
has indeed two solutions which furthermore are independent of the ST constants. These solutions satisfy $q_{+2}q_{-2}=-1$, with

\begin{equation}
q_{-2}=\frac{\delta \omega ^{\text{(4)}}_{-2}\pm i \sqrt{\left(\delta \omega ^{\text{(3)}}_{-2}\right)^2-\left(\delta \omega ^{\text{(4)}}_{-2}\right)^2}}{\delta \omega ^{\text{(3)}}_{-2}}\, ,
\end{equation}
resulting in a total shift in the frequency

\begin{equation}\label{deltaomegaoddapp}
\delta\omega_{s}=\delta\omega_{s}^{*}=\pm\frac{16 \sqrt{\left(\delta \omega ^{\text{(3)}}_{-2}\right)^2-\left(\delta \omega ^{\text{(4)}}_{-2}\right)^2}}{K_2}\equiv \pm \delta\omega_{\rm break}\, .
\end{equation}
We remark that the structure of these coefficients is crucial in order to observe these properties. Additionally, our numerical results also indicate that $\delta\omega_{s}^{(2)}=0$. Although we do not know the origin of this result, assuming $\delta\omega_{s}^{(2)}=0$  leads to an interesting consequence. If we use this condition in the relations \req{eq:relationsparityodd} above, we find that $\delta\omega_{-2}^{(4)}=-i \delta\omega_{-2}^{(3)}/(2K_1)$, obtaining that the $q_{-2}$ parameter is given by 

\begin{equation}
q_{-2}=\frac{2 i K_1}{1\pm \sqrt{4 K_1^2+1}}\, ,
\end{equation}
and thus implying that the polarization of QNMs in these parity-violating theories is actually theory-independent.

In principle it would be possible to obtain the relations \req{eq:relationsparityeven} and \req{eq:relationsparityodd} directly from the form of the equations, \textit{e.g.}, by observing that the potential itself \req{eq:deltaVCq} satisfies the same relations. However, the potential may not show these relations explicitly, since it depends on the choice of variables. It should always be possible, via changes of variables, to write the potential in a way that these relations are explicit, but we find this to be a very difficult and not useful exercise in the rotating case.  The frequencies on the other hand always exhibit these relations since they are independent of changes of variables. 
In practice, we compute the coefficients $\delta\omega_{s}^{(j)}$, $\delta\omega_{s}^{*(j)}$ numerically and check that the relations \req{eq:relationsparityeven} and \req{eq:relationsparityodd} above are satisfied up to terms of order $\mathcal{O}(\chi^{n+1})$ on account of the truncation in the spin expansion. The fact that the relations are not satisfied exactly introduces a mild dependence on $C_{\pm 2}$ in the correction to the QNM frequencies \req{deltaomegaevenapp}  and \req{deltaomegaoddapp} which we use to estimate our error, as explained in the main text. 

Finally, if the action contains both parity-preserving and parity-violating terms, then the coefficients $\delta\omega_{s}^{(j)}$, $\delta\omega_{s}^{*(j)}$ are a linear combination of \req{eq:relationsparityeven} and \req{eq:relationsparityodd}. Then, one can again solve the equations \req{eq:equationsomegaapp} and check that they admit for two solutions with associated frequencies given by 

\begin{equation}\label{eq:combinationrule}
\delta\omega_{\rm total}^{\pm}=\frac{\delta\omega^{+}+\delta\omega^{-}}{2}\pm  \sqrt{\frac{\left(\delta\omega^{+}-\delta\omega^{-}\right)^2}{4}+\delta\omega_{\rm break}^2}\, ,
\end{equation}
which is \req{eq:combinationrule} in the main text.  In addition, if we use that $\delta\omega_{s}^{(2)}=0$ for the parity-breaking theory, we can find a closed formula for the polarization parameters $q_{s}$ for each of these modes, 

\begin{align}
q_{s}^{\pm}&=\frac{\pm\sqrt{4 K_1^2+1} \sqrt{\left(\delta \omega ^+-\delta \omega ^-\right)^2+4 \delta \omega _{\rm{break}}^2}-s \delta \omega _{\rm{break}}}{\sqrt{4 K_1^2+1} \left(\delta \omega ^+-\delta \omega ^-\right)+4 i K_1 \delta \omega _{\rm{break}}}\, ,
\end{align}
with $s=\pm2$.

\bibliographystyle{apsrev4-1} 
\vspace{1cm}
\bibliography{Gravities} 

\end{document}